\documentclass{iopart}
\usepackage{iopams,bm,mathrsfs}
\usepackage[mathcal]{euscript}
\usepackage{temp}
\usepackage{xcolor}
\usepackage[breaklinks=true,unicode=true,urlcolor = blue,colorlinks = true,citecolor = blue,linkcolor = blue]{hyperref}
\usepackage{graphicx}
\usepackage[numbers,sort&compress]{natbib}
\renewcommand{\vec}[1]{\bm{#1}}
\renewcommand{\imath}{\rmi}
\begin{document}

\title{Magnetization  in narrow  ribbons: curvature effects}

\author{Yuri~Gaididei$^1$, Arseni~Goussev$^2$, Volodymyr~P.~Kravchuk$^1$, Oleksandr~V.~Pylypovskyi$^3$, J.~M.~Robbins$^4$, Denis~D.~Sheka$^3$, Valeriy~Slastikov$^4$, and Sergiy~Vasylkevych$^4$}

\address{$^1$ Bogolyubov Institute for Theoretical Physics of National Academy of Sciences of Ukraine, 03680 Kyiv, Ukraine}
\address{$^2$ Department of Mathematics, Physics and Electrical Engineering, Northumbria University, Newcastle upon Tyne, NE1 8ST, UK}
\address{$^3$ Taras Shevchenko National University of Kyiv, 01601 Kyiv, Ukraine}
\address{$^4$ School of Mathematics, University of Bristol, Bristol, BS8 1TW, UK}
\eads{
\mailto{ybg@bitp.kiev.ua}, 
\mailto{arseni.goussev@northumbria.ac.uk}, 
\mailto{vkravchuk@bitp.kiev.ua}, 
\mailto{engraver@knu.ua}, 
\mailto{J.Robbins@bristol.ac.uk}, 
\mailto{sheka@univ.net.ua}, 
\mailto{s.vasylkevych@bristol.ac.uk}
}

%
%
%
%
\begin{abstract}
	
A ribbon is a surface swept out by a line segment turning as it moves along a central curve. For narrow magnetic ribbons, for which the length of the  line segment is much less than the length of the  curve, the anisotropy induced by the magnetostatic interaction is biaxial, with hard axis normal to the ribbon and easy axis along the central curve. The micromagnetic energy of a narrow ribbon reduces to that of a one-dimensional ferromagnetic wire, but with curvature, torsion and local anisotropy modified by the rate of turning.  These general results are applied to two examples, namely  a helicoid ribbon, for which the central curve is a straight line, and a M\"obius ribbon, for which the central curve is a circle about which the line segment executes a $180^\circ$ twist. In both examples, for large positive tangential anisotropy, the ground state magnetization lies tangent to the central curve.  As the tangential anisotropy is decreased, the ground state magnetization undergoes a transition, acquiring an in-surface component perpendicular to the central curve.  For the helicoid ribbon, the transition occurs at vanishing anisotropy, below which the  ground state is uniformly perpendicular to the central curve.  The transition for the M\"obius ribbon is more subtle; it  occurs at a positive critical value of the anisotropy, below which the ground state is nonuniform.  For the helicoid ribbon, the dispersion law for spin wave excitations about the tangential state  is found to exhibit an asymmetry determined by the geometric and magnetic chiralities.
\end{abstract}
\submitto{\JPA}
%
%
\maketitle
%
%

\section*{Introduction}
\label{sec:intro}
The emerging area of {\it magnetism in curved geometries} encompasses a range of fascinating geometry-induced effects in the magnetic properties of materials \cite{Streubel16a}.
Theoretical investigations in this area are providing new insights into the behaviour of curved magnetic nanostructures and the control of their magnetic excitations, with applications to shapeable magnetoelectronics \cite{Makarov16} and prospective 
energy-efficient data storage, among others. 

In continuum models, the magnetization is represented by  a three-dimensional unit-vector field $\vec{m}(\vec{r})$. The study of  curvature--induced effects in vector-field models 
in one- and two-dimensional geometries has a rather long history \cite{Kamien02,Nelson04,Bowick09,Turner10}. In spite of numerous results 
\cite{Kamien02,Nelson04,Bowick09,Turner10}, the problem is far from being fully solved. In the majority of these  studies, the vector field is taken to be tangent to the domain. In particular, a general expression for the surface energy of a tangential director field describing a nematic liquid crystal  in a curvilinear shell was recently obtained \cite{Napoli12,Napoli12a,Napoli13,Napoli13a}, with possible applications using different geometries and orientational ordering  \cite{Segatti14,Manyuhina14,Oliveira16}. The assumption of a strictly tangential field was also used in a study of the role of curvature in the interaction between defects in 2D $XY$-like models, with applications to superfluids, superconductors, and liquid crystals deposited on curved surfaces \cite{Vitelli04}.

Very recently a fully 3D approach was developed for thin magnetic shells and wires of arbitrary shape \cite{Gaididei14, Sheka15}. This approach yields an energy for arbitrary curves and surfaces and for arbitrary magnetization fields under the assumption that the anisotropy  greatly exceeds the dipolar interaction, so that
\begin{equation} \label{eq:enegry-general}
E = \int\!\! \mathrm{d} \vec{r} \left(\mathscr{E}_{\mathrm{ex}} +\mathscr{E}_{\mathrm{an}}\right).
\end{equation} 
Here $\mathscr{E}_{\mathrm{ex}}$  is the exchange energy density and $\mathscr{E}_{\mathrm{an}}$ is the density of effective anisotropy interaction. We consider the model of isotropic exchange, $\mathscr{E}_{\mathrm{ex}} =  \left( \vec{\nabla} m_i \right)\cdot \left( \vec{\nabla} m_i \right)$, where $m_i$ with $i=1,2,3$ describes the cartesian components of magnetization. Therefore in cartesian coordinates, the sample geometry appears  only through the anisotropy term via  the spatial variation of the anisotropy axis; for example, in the case of a uniaxial curved magnet, $\mathscr{E}_{\mathrm{an}}$ is given by  $K \left(\vec{m}\cdot \vec{e}_A\right)^2$, where the unit vector $\vec{e}_A = \vec{e}_A(\vec{r})$ determines the direction of the easy axis.

In curvilinear coordinates adapted to the sample geometry, the spatial variation of the anisotropy axes is automatically accounted for, and the anisotropy energy density assumes its usual translation-invariant form.  Instead, the exchange energy acquires two additional terms, which describe contributions to $\left( \vec{\nabla} m_i \right)\cdot \left( \vec{\nabla} m_i \right)$ due to the spatial variation of the coordinate frame \cite{Sheka15}, namely curvilinear-geometry-induced effective anisotropy and curvilinear-geometry-induced effective Dzyaloshinskii--Moriya interaction.  For magnetic shells, these contributions may be expressed in terms of  local curvatures \cite{Gaididei14}; for magnetic wires, in terms of curvature and torsion \cite{Sheka15}. Below we review briefly some manifestations of these contributions, which have been reported elsewhere.

(i) \emph{Curvilinear-geometry-induced effective anisotropy}. Geometry-induced anisotropy can have a significant effect on the ground-state magnetization profile, rendering it no longer strictly tangential, even in the case of strong easy-tangential anisotropy. For example, for a helical nanowire with strong anisotropy directed along the wire, the ground-state magnetization is always tilted in the local rectifying surface, with tilting angle dependent on the product of the curvature and the torsion \cite{Sheka15c,Pylypovskyi15e}. For two-dimensional geometries with nontrivial topology,  a  striking manifestation of geometry-induced anisotropy is  shape--induced patterning.  In spherical shells, a strictly in-surface magnetization is forbidden due to the hairy--ball theorem \cite{Milnor78}. Instead, the ground-state magnetization profile has two oppositely disposed vortices \cite{Kravchuk12a}. Another nontrivial example is the M{\"o}bius ring. Since a M\"obius ring is a nonorientable surface, its topology  forces a discontinuity in any nonvanishing normal vector field. Recently we proposed  that magnetic nanostructures shaped as M\"{o}bius strips possess non-volatility in their magneto-electric response due to the presence of topologically protected magnetic domain walls in materials with an out-of-plane orientation of the easy axis of magnetization \cite{Pylypovskyi15b}. In  both of these  examples, the link between surface topology and  magnetization is a consequence of geometry--dependent anisotropy. 

(ii) \emph{Curvilinear-geometry-induced effective Dzyaloshinskii--Moriya interaction}. Recently, the role of  curvature in  domain wall pinning was elucidated \cite{Yershov15b}; a local bend in a nanowire is the source of a pinning potential for transversal domain walls. Chiral symmetry-breaking due to a geometry-induced  Dzyaloshinskii--Moriya interaction strongly impacts the domain wall dynamics and allows domain wall motion under the action of different spin--torques, e.g. field--like torques \cite{Pylypovskyi15e} and anti--damping torques \cite{Yershov15c}. In the particular case of a helical nanowire,  torsion can produce negative domain wall mobility \cite{Pylypovskyi15e,Yershov15c}, while  curvature can produce a shift in the Walker breakdown \cite{Yershov15c}.

We have briefly described a theoretical framework for studying different curvilinear systems, including 1D nanowires and 2D nanoshells. In this approach we suppose that the
effects of nonlocal dipole-dipole interactions can be reduced to an effective easy-surface anisotropy. In the 1D case, this reduction has been rigorously justified in the limit where the diameter of the wire $h$ is much smaller than its length $L$ \cite{Slastikov12}. Similar arguments have been provided in the 2D case for planar thin films \cite{Kohn05} and thin shells \cite{Slastikov05} where the surface thickness $h$ is much less than the lateral size $L$. 

In the current study we consider a \emph{ribbon}, which represents a curve with an infinitesimal neighbourhood of a surface along it \cite{Sternberg12}. For a narrow ribbon whose thickness $h$ is much less than its width $w$, which in turn is much less than its length $L$, namely $h\ll w\ll L$, another micromagnetic limit is realized. We show that the micromagnetic energy can be reduced to the energy of a wire with modified curvature, torsion and anisotropy. We illustrate this approach with two examples, namely a narrow helicoid ribbon and a M\"obius ribbon.  The existence of a new nonhomogeneous ground state is predicted for the M\"obius ribbon over a range of anisotropy parameter $K$. The prediction is  confirmed by full scale spin--lattice simulations. We also analyse the magnon spectrum for a narrow helicoid ribbon: unlike the magnon spectrum for a straight wire, there appears an asymmetry in the dispersion law caused by the geometric and magnetic chiralities.

The paper is organized as follows. In Section~\ref{sec:model} we derive the micromagnetic energy for a narrow ribbon, which may be interpreted as a modification of the 1D micromagnetic energy of its central curve. We illustrate the model by two examples,  a helicoid ribbon (Section ~\ref{sec:helicoid-ribbon}) and a M\"obius ribbon 
(Section~\ref{sec:Moebius-ribbon}).   Concluding remarks are given in Section~\ref{sec:conclusion}.
The justification of the magnetostatic energy for ribbons and strips is presented in \ref{app:rigorous}. The spin-lattice simulations are detailed in \ref{app:sim}.


\section{Model of narrow ribbon vs thin wire}
\label{sec:model}

\subsection{Thin ferromagnetic wire}
\label{sec:wire}

Here we consider a ferromagnetic wire described by a curve $\vec{\gamma}(s)$  with fixed cross-section of area $S$, parameterized by arc length $s \in [0,L]$, where  $L$ is  the length of the  wire. It has been shown  \cite{Slastikov12} that the  properties of sufficiently  thin ferromagnetic wires 
of circular (or square) cross section are described by  a reduced one-dimensional energy  given by a sum of exchange and local anisotropy terms,
\begin{equation} \label{eq:En-wire}
\eqalign{
E^{\mathrm{wire}} &= 4\pi M_s^2 S \! \int\limits_0^L \!\! \mathrm{d}s \bigl( \mathscr{E}_{\mathrm{ex}}^{\mathrm{wire}} + \mathscr{E}_{\mathrm{an}}^{\mathrm{wire}} \bigr),\\
\mathscr{E}_{\mathrm{ex}}^{\mathrm{wire}} &= \ell^2|\vec{m}'|^2
\qquad \mathscr{E}_{\mathrm{an}}^{\mathrm{wire}}  = -\frac{Q_1}{2}\left(\vec{m}\cdot\vec{e}_{\scriptsize \textsc{t}}\right)^2.
}
\end{equation}
Here, $\vec{m}(s)$ denotes the unit magnetization vector, prime $'$ denotes  derivative with respect to $s$, $M_s$ is the saturation magnetization, and $\ell=\sqrt{\mathcal{A}/ 4\pi M_s^2}$ is the exchange length with $\mathcal{A}$ being the exchange constant. The local anisotropy is uniaxial, with easy axis along the tangent  $\vec{e}_{\scriptsize \textsc{t}} = \vec{\gamma}'$.  The normalized anisotropy constant (or quality factor) $Q_1$ incorporates the intrinsic crystalline anisotropy $K_1$ as well as a geometry-induced magnetostatic contribution, 
\begin{equation} \label{eq:Q_1}
Q_1= \frac{K_1}{2\pi M_s^2}+\frac{1}{2}.
\end{equation}
Note that the  shape-induced biaxial anisotropy is caused by the asymmetry of the  cross-section. In particular, for a rectangular cross-section, the anisotropy coefficients are determined by Eq.~\eqref{eq:en-rib}; for elliptical cross-sections, see \cite{Slastikov12}.

It is convenient to express the magnetization in terms of  the Frenet-Serret frame comprised of the tangent  $\vec{e}_{\scriptsize \textsc{t}}$, the normal
$\vec{e}_{\scriptsize \textsc{n}} = \vec{e}_{\scriptsize \textsc{t}}'/ |\vec{e}_{\scriptsize \textsc{t}}'|$, and the binormal $\vec{e}_{\scriptsize \textsc{b}}= \vec{e}_{\scriptsize \textsc{t}} \times \vec{e}_{\scriptsize \textsc{n}}$.  These satisfy the Frenet-Serret equations,
\begin{equation} \label{eq:Frenet-Serret}
\vec{e}_\alpha ' = F_{\alpha \beta }\vec{e}_\beta , \qquad \left\|F_{\alpha \beta } \right\| =
\left(
\begin{array}{ccc}
0 & \kappa & 0 \\
-\kappa & 0 & \tau  \\
0 &- \tau  & 0
\end{array}
\right),
\end{equation}
where $\kappa(s)$ and $\tau(s)$ are the curvature and torsion of $\vec{\gamma}(s)$, respectively. 
Letting
\begin{equation*} 
\vec{m} = \sin\varTheta \cos\varPhi \, \vec{e}_{\scriptsize \textsc{t}} + \sin\varTheta \sin\varPhi \,\vec{e}_{\scriptsize \textsc{n}} + \cos \varTheta \, \vec{e}_{\scriptsize \textsc{b}},
\end{equation*}
where $\varTheta$ and $\varPhi$ are functions of $s$ (and time $t$, if dynamics is considered),
one can show \cite{Sheka15}  that the exchange and anisotropy energy densities are given by
\begin{equation*} 
\eqalign{
\mathscr{E}_{\mathrm{ex}}^{\mathrm{wire}} &= \ell^2\left[\varTheta' - \tau\sin\varPhi\right]^2 + \ell^2\left[\sin\varTheta(\varPhi' + \kappa) - \tau \cos\varTheta \cos\varPhi\right]^2, 
\\
\mathscr{E}_{\mathrm{an}}^{\mathrm{wire}}  &= -\frac{Q_1}{2} \sin^2\varTheta\cos^2\varPhi.}
\end{equation*}


\subsection{Narrow ferromagnetic ribbon}
\label{sec:ribbon}

As above, let $\vec{\gamma}(s)$ denote a three-dimensional curve parametrized by arc length.
Following \cite{Sternberg12}, we take a \emph{ribbon} to be a two-dimensional surface swept out by a line segment centred at and perpendicular to $\gamma$,  moving (and possibly turning) along $\gamma$ The ribbon may parametrized as 
\begin{equation} \label{eq:surface}
\vec{\varsigma}(s,v) = \vec{\gamma}(s) + v\cos\alpha(s)\vec{e}_{\scriptsize \textsc{n}} + v\sin\alpha(s)\vec{e}_{\scriptsize \textsc{b}}, \quad v\in\left[-\frac{w}{2},\frac{w}{2}\right],
\end{equation}
where $w$ is the width of the segment (assumed to be small enough so that $\vec{\varsigma}$ has no self-intersections) and $\alpha(s)$ determines the orientation of the segment with respect to the normal and binormal. We  construct a three-frame $\left\{\vec{e}_1,\vec{e}_2, \vec{e}_3\right\}$ on the ribbon given by 
\begin{equation} \label{eq:2D-basis}
\vec{e}_\mu = \frac{\partial_\mu \vec{\varsigma}}{|\partial_\mu \vec{\varsigma}|},\ \ \mu = 1, 2, \quad
\vec{e}_3 = \vec{e}_1\times \vec{e}_2.
\end{equation}
Here and in what follows, we use Greek letters $\mu, \nu, {\rm etc} = 1,2 $ to denote indices restricted to the ribbon surface. Using the Frenet--Serret equations \eqref{eq:Frenet-Serret}, one can show  that \eqref{eq:2D-basis} constitute an orthonormal frame, with $\vec{e}_1$ and $\vec{e}_2$ tangent to the ribbon and $\vec{e}_3$ normal to it. It follows that the first fundamental form (or metric), $g_{\mu \nu } = \partial_\mu \vec{\varsigma} \cdot \partial_\nu \vec{\varsigma}$, is diagonal. The second fundamental form, $b_{\mu \nu }$, is given by $b_{\mu \nu } = \vec{e}_3\cdot \partial^2_{\mu,\nu} \vec{\varsigma}$. The Gau\ss{} and mean curvatures are given respectively by the determinant and trace of $|| H_{\mu \nu} || =  ||b_{\mu \nu }/\sqrt{g_{\mu \mu }g_{\nu \nu }}||$.

We consider a thin ferromagnetic shell about the ribbon of thickness $h$, where 
\begin{equation} \label{eq:thin}
h \ll w, L.
\end{equation}
The shell is comprised of points $\vec{\varsigma}(s,v) + u \vec{e}_3$, where $u \in [-h/2,h/2]$. We express the unit magnetization inside the shell in terms of the frame $\vec{e}_\alpha$ as 
\begin{equation*} 
\vec{m} = \sin\theta \cos\phi \, {\vec{e}}_1 + \sin\theta \sin\phi \,{\vec{e}}_2 + \cos \theta \, {\vec{e}}_3,
\end{equation*} 
where $\theta$ and $\phi$ are functions of the surface coordinates $s,v$ (and time $t$, for dynamical problems), but are independent of the transverse coordinate $u$. The micromagnetic energy of a thin shell reads
\begin{subequations} \label{eq:E-shell}
\begin{equation} \label{eq:E-shell-total}
E^{\mathrm{shell}} = 4\pi M_s^2 h \int\limits_0^L \! \mathrm{d}s \!\! \int\limits_{-w/2}^{w/2}\! \!\! \sqrt{g} \mathrm{d}v \left( \mathscr{E}_\mathrm{ex}^{\mathrm{shell}} + \mathscr{E}_\mathrm{an}^{\mathrm{shell}} \right) + E_{\mathrm{ms}}^{\mathrm{shell}},
\end{equation}
where $g=\det\left(g_{\mu \nu}\right)$. The exchange energy density in \eqref{eq:E-shell-total} is given by \cite{Gaididei14,Sheka15}
\begin{equation} \label{eq:E-shell-exchange}
\mathscr{E}_{\mathrm{ex}}^{\mathrm{shell}} = \ell^2\left[\vec{\nabla}\theta  - \vec{\varGamma }(\phi )\right]^2 + \ell^2 \left[\sin\theta \left(\vec{\nabla}\!\phi-\vec{\varOmega }\right) - \cos \theta \frac{\partial \vec{\varGamma }(\phi )}{\partial\phi }\right]^2\!\!,
\end{equation}
where $\vec{\nabla} \equiv \vec{e}_\mu \nabla_\mu$ denotes a surface del operator in its curvilinear form with components $\nabla_\mu  \equiv \left(g_{\mu\mu}\right)^{-1/2} \partial_\mu$, the vector $\vec{\varOmega}$ is a spin connection with components $\varOmega_\mu = \vec{e}_1\cdot \nabla_\mu  \vec{e}_2$, and the vector $\vec{\varGamma}(\phi)$ is given by $||H_{\mu\nu}|| \left(\begin{array}{c} \cos\phi\\\sin\phi\end{array}\right)$. The next term in the energy functional, $\mathscr{E}_\mathrm{an}^{\mathrm{shell}}$, is the anisotropy energy density of the shell:
\begin{equation} \label{eq:Q1-Q3-shell}
\mathscr{E}_{\mathrm{an}}^{\mathrm{shell}}  = -\frac{K_1}{4\pi M_s^2} \left(\vec{m}\cdot\vec{e}_1\right)^2 - \frac{K_3}{4\pi M_s^2} \left(\vec{m}\cdot\vec{e}_3\right)^2,
\end{equation} 
\end{subequations}
where $K_1$  and $K_3$ are the tangential and normal anisotropy coefficients of the intrinsic crystalline anisotropy. The magnetostatic energy, $E_{\mathrm{ms}}^{\mathrm{shell}}$, has, in the general case, a nonlocal form. The local form is restored in the limit  of thin films \cite{Gioia97,Carbou01,Kohn05a} and thin shells \cite{Slastikov05,Fratta16}.

We proceed to consider the narrow-ribbon limit,
\begin{equation} \label{eq:narrow-defn}
\frac{w^2}{\ell} \leq h \ll w \ll \ell \lesssim  L.
\end{equation}
Keeping  leading-order terms in $w/L$ we obtain that the geometrical properties of ribbon are determined by
\begin{equation*} 
\left\| g_{\mu \nu}^{\mathrm{ribbon}} \right\| = \mathrm{diag}(1,1), \quad \left\| H_{\mu \nu}^{\mathrm{ribbon}}\right\| = \left(
\begin{array}{cc}
-\kappa \sin\alpha 	& \alpha'+\tau \\
\alpha'+\tau		&	0
\end{array}
\right),
\end{equation*}
In the same way, we obtain from \eqref{eq:E-shell} the following:
\begin{subequations} \label{eq:E-ribbon}
\begin{equation} \label{eq:E-ribbon-1}
\eqalign{
& E^{\mathrm{ribbon}} = 4\pi M_s^2 hw \int \mathrm{d}s\, \left( \mathscr{E}_\mathrm{ex}^{\mathrm{eff}} + \mathscr{E}_\mathrm{an}^{\mathrm{eff}}
\right),\\	
&\mathscr{E}_\mathrm{ex}^{\mathrm{eff}}=\ell^2 \left(\theta' -\varGamma_1\right)^2 + \ell^2 \left[\sin\theta \left(\phi'-\varOmega_1\right)\! - \!\cos \theta \frac{\partial\Gamma_1}{\partial\phi }\right]^2,\\
&\mathscr{E}_\mathrm{an}^{\mathrm{eff}}=\ell^2\varGamma_2^2 + \ell^2 \cos^2  \theta \left( \frac{\partial\varGamma_2}{\partial\phi }\right)^2  + \mathscr{E}_{\mathrm{an}}^{\mathrm{ribbon}},
}
\end{equation}
where the effective spin connection $\varOmega_1$ and vector $\vec{\varGamma}$ are given by
\begin{equation} \label{eq:ribbon-wire-1}
\fl 
\varOmega_1 = -\kappa \cos\alpha, \quad \varGamma_1 = -\kappa \sin\alpha \cos\phi + (\alpha' + \tau)\sin\phi, \quad \varGamma_2 = (\alpha' + \tau) \cos\phi.
\end{equation}
The last term in the energy density, $\mathscr{E}_{\mathrm{an}}^{\mathrm{ribbon}}$, is the effective anisotropy energy density of the narrow ribbon. Using arguments similar to those in  \cite{Gioia97,Kohn05a,Slastikov05}, it can be shown that
\begin{equation} \label{eq:Q1-Q3ribbon-1}
\mathscr{E}_{\mathrm{an}}^{\mathrm{ribbon}}  = -\frac{Q_1}{2}\left(\vec{m}\cdot\vec{e}_1\right)^2 - \frac{Q_3}{2}\left(\vec{m}\cdot\vec{e}_3\right)^2.
\end{equation} 
Here $Q_1$ and $Q_3$ incorporate the  intrinsic crystalline anisotropies $K_1$  and $K_3$ as well as geometry-induced magnetostatic contributions:
\begin{equation} \label{eq:Q1-Q3ribbon-2}
\fl %
Q_1 = \frac{K_1}{2\pi M_s^2} + Q_r, \qquad Q_r = \frac{h}{\pi w}\ln\frac{w}{h}, \qquad
Q_3 = -1 + \frac{K_3}{2\pi M_s^2} +2 Q_r,
\end{equation} 
\end{subequations}
see the justification in \ref{app:rigorous}. In the particular case of soft magnetic materials, where $K_1 = K_3 = 0$, the anisotropy $\mathscr{E}_{\mathrm{an}}^{\mathrm{ribbon}}$ is due entirely to the magnetostatic interaction.  From \eqref{eq:Q1-Q3ribbon-2},  we get $Q_1 = Q_r\ll1$ and $Q_3 = -1+2Q_r$.

The induced anisotropy is biaxial, with easy axis along the central curve as for a thin wire (cf \eqref{eq:Q_1})  and  hard axis normal to the surface as for a thin shell. Indeed, one can recast the narrow-ribbon energy  \eqref{eq:E-ribbon} in the form of the thin-wire energy \eqref{eq:En-wire} with biaxial anisotropy, as follows:
\begin{equation} \label{eq:narrow-ribbon-wire-form}
\fl
\eqalign{
\mathscr{E}_\mathrm{ex}^{\mathrm{eff}} &= 
\ell^2 \left(\theta' -\tau^{\mathrm{eff}} \sin\Psi\right)^2 + \ell^2 \Bigl[\sin\theta \left(\Psi' + \kappa^{\mathrm{eff}}\right)- \tau^{\mathrm{eff}} \cos \theta \cos\Psi \Bigr]^2,\\
\mathscr{E}_\mathrm{an}^{\mathrm{eff}} &= -\frac{Q^{\mathrm{eff}}_1}{2}\sin^2\theta\cos^2\phi - \frac{Q^{\mathrm{eff}}_3}{2}\cos^2\theta.}
\end{equation}
In \eqref{eq:narrow-ribbon-wire-form}, the effective curvature and torsion are given by
\begin{equation} \label{eq:kappa-tau}
\kappa^{\mathrm{eff}} = \kappa\cos\alpha - \beta', \qquad \tau^{\mathrm{eff}} = \sqrt{\kappa^2\sin^2\beta + (\alpha'+\tau)^2},
\end{equation}
the angle $\Psi$ is defined by
\begin{equation*} 
\Psi = \phi + \beta, \qquad \tan \beta = -\frac{\kappa \sin\alpha}{\alpha'+\tau},
\end{equation*}
and the effective anisotropies are given by
\begin{equation} \label{eq:narrow-ribbon-Qeff}
Q^{\mathrm{eff}}_1 = Q_1 - 2\ell^2 \left(\alpha' + \tau\right)^2, \quad Q^{\mathrm{eff}}_3 = Q_3 - 2\ell^2 \left(\alpha' + \tau\right)^2.
\end{equation}

	
\section{Helicoid ribbon}
\label{sec:helicoid-ribbon}

\begin{figure}
\begin{center}
\includegraphics[width=\linewidth]{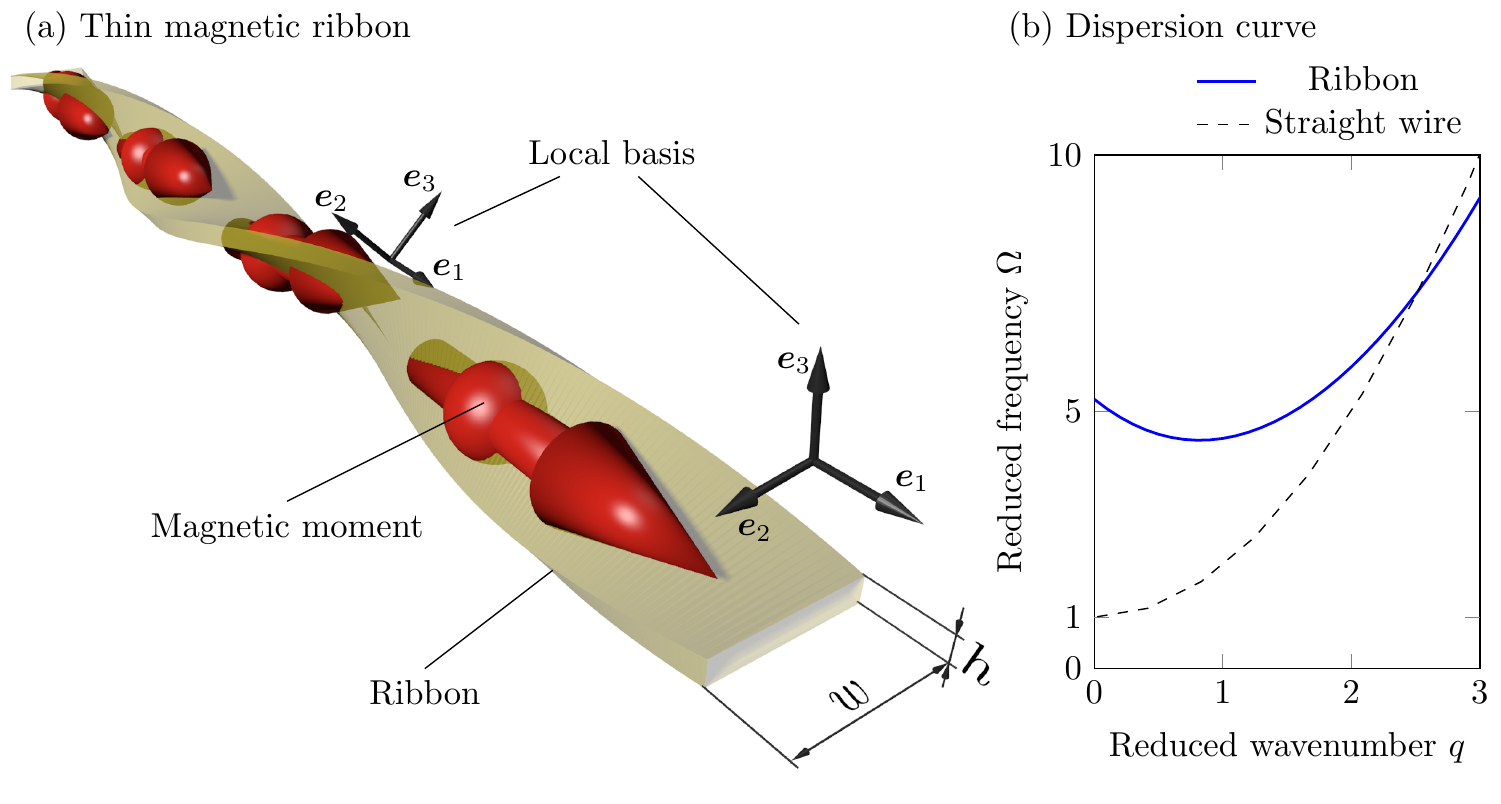}
\end{center}
\caption{\textbf{Magnetic helicoid ribbon:} (a)~A sketch of the ribbon. 
(b)~Dispersion curve according to Eq.~\eqref{eq:disp} (solid blue line) in comparison with the dispersion of the straight wire $\Omega_{\mbox{str}}=1+q^2$.
}
\label{fig:helicoid}
\end{figure}

The helicoid ribbon has a straight line, which has vanishing curvature and torsion, as its central curve.  We take $\vec{\gamma}(s) = s\, \hat{\vec{z}}$. The  rate of turning about $\vec{\gamma}$ is constant, and we take  $\alpha(s) = \mathcal{C} s/s_0$, where  
the chirality $\mathcal{C}$ is $+1$ for a right-handed helicoid and $-1$ for a left-handed  helicoid. From \eqref{eq:surface}, the parametrized surface is given by
\begin{equation*} 
\vec{\varsigma}(s,v) = \hat{\vec{x}}\, v \cos \left(\frac{s}{s_0}\right) + \hat{\vec{y}}\, \mathcal{C} v \sin \left(\frac{s}{s_0}\right) + \hat{\vec{z}}\, s, \quad v\in\left[-\frac{w}{2},\frac{w}{2}\right].
\end{equation*}
The boundary curves, given by $\vec{\varsigma}(s,\pm w/2)$, are helices, see Fig.~\ref{fig:helicoid}~(a). It is well known that the curvature and torsion essentially influence the spin-wave dynamics in a helix wire, acting as an effective magnetic field \cite{Sheka15c}. One can expect similar behaviour in a helicoid ribbon.

From \eqref{eq:kappa-tau} and \eqref{eq:narrow-ribbon-Qeff}, the effective curvature, torsion and anisotropies  are given by 
 \begin{equation} \label{eq:kappa-tau-helicoid}
\fl
\kappa^{\mathrm{eff}} = 0, \quad \tau^{\mathrm{eff}} = \frac{\mathcal{C}}{s_0}, \quad Q^\mathrm{eff}_1 =Q_1 - 2 \left(\frac{\ell}{s_0}\right)^2, \quad Q^\mathrm{eff}_3= Q_3 - 2 \left(\frac{\ell}{s_0}\right)^2.
\end{equation}
From  \eqref{eq:narrow-ribbon-wire-form}, the energy density is given by
\begin{equation} \label{eq:narrow-helicoid-energy}
\eqalign{
\mathscr{E}_\mathrm{ex}^{\mathrm{eff}} &= \frac{\ell^2}{s_0^2} \left[ (s_0\theta'- \mathcal{C}\sin\phi)^2 + (s_0 \sin\theta \phi' - \mathcal{C} \cos\theta\cos\phi)^2 \right] ,\\
 \mathscr{E}_\mathrm{an}^{\mathrm{eff}} &= 
 - \frac{Q^{\mathrm{eff}}_1}{2}\sin^2\theta\cos^2\phi - \frac{Q^{\mathrm{eff}}_3}{2}\cos^2\theta.}
\end{equation}

Let us consider the particular case  of soft magnetic materials ($K_1 = K_3 = 0$).  Under the reasonable assumption $\ell \ll s_0$, we see that $Q_3\approx -1$, so that the easy-surface anisotropy dominates the energy density and acts as an in-surface constraint.  Taking $\theta = \pi/2$ to accommodate this constraint, we obtain the (further) reduced energy density
\begin{equation*} 
\mathscr{E}^{\mathrm{eff}} = \ell^2 {\phi'}^2  - \frac{Q_1}{2}\cos^2\phi,
\end{equation*}
which depends only on the in-surface orientation $\phi$.  The ground states have $\phi$ constant, with orientation depending on the sign of the tangential-axis anisotropy  $Q_1$. For $Q_1 > 0$,  the ground states are
\begin{equation} \label{eq:helicoid-gs}
\theta^{\scriptsize \textsc{t}}=\frac{\pi}{2},\qquad \cos \phi^{\scriptsize \textsc{t}}  = \mathfrak{C},
\end{equation}
where the magnetochirality  $\mathfrak{C}=\pm1$ determines whether the magnetisation $\vec{m}$ is parallel ($\mathfrak{C}=1$) or antiparallel ($\mathfrak{C}=-1$) to the helicoid axis.  For $Q_1 < 0$, the ground states are given by 
\begin{equation*} 
\theta^{\scriptsize \textsc{n}}=\frac{\pi}{2},\qquad \phi^{\scriptsize \textsc{n}}  = \mathfrak{C}  \frac{\pi}{2},
\end{equation*}
where the magnetochirality  $\mathfrak{C}=\pm1$ determines whether the magnetisation $\vec{m}$ is parallel ($\mathfrak{C}=1$) or antiparallel ($\mathfrak{C}=-1$) to the normal $\vec{e}_{\scriptsize \textsc{n}}$. This behaviour is similar to that of a ferromagnetic helical wire, which was recently studied in Ref.~\cite{Sheka15c}.

\subsection{Spin-wave spectrum in a helicoid ribbon}
\label{sec:helicoid-spin-waves}

Let us consider spin waves  in a helicoid ribbon on the tangential ground state \eqref{eq:helicoid-gs}.  We write 
\begin{equation*} 
\theta = \theta^{\scriptsize \textsc{t}} + \vartheta(\chi,\tilde{t}), \quad \phi= \phi^{\scriptsize \textsc{t}} + \varphi(\chi,\tilde{t}), \qquad |\vartheta|, |\varphi| \ll 1,
\end{equation*}
where $\chi=s/s_0$ and $\tilde{t}=\Omega_0 t$ with $\Omega_0 = (2\gamma_0/M_s)(\ell/s_0)^2$. Expanding the energy density \eqref{eq:narrow-helicoid-energy} to quadratic order in the $\vartheta$ and $\varphi$, we obtain 
\begin{equation*} 
\eqalign{
\mathscr{E}&= \left(\frac{\ell}{s_0}\right)^2 \left[ \left(\partial_\chi\vartheta\right)^2 + \left(\partial_\chi\varphi\right)^2\right] + 2\mathcal{C}\mathfrak{C} \left(\frac{\ell}{s_0}\right)^2  \left(\vartheta \partial_\chi\varphi - \varphi \partial_\chi\vartheta\right)\\
& + \left[Q_1-Q_3 +2\left(\frac{\ell}{s_0}\right)^2\right]\frac{\vartheta^2}{2}  + Q_1 \frac{\varphi^2}{2}.
}
\end{equation*}
The linearised Landau--Lifshits equations have the form of a generalized Schr{\"o}dinger equation for the complex-valued function $\psi =\vartheta +i\varphi$ \cite{Sheka15c},
\begin{equation} \label{eq:Schroedinger}
-i\partial_{\tilde{t}} \psi  = H \psi  + W\psi ^*, \quad H = \left(-i\partial_\chi  -A \right)^2 +U,
\end{equation}
where the ``potentials'' have the following form: 
\begin{equation} \label{eq:potentials-helicoid}
\fl
U =  - \frac{1}{2} + \frac{1}{4} \left(\frac{s_0}{\ell}\right)^2 \left(2Q_1-Q_3 \right), \qquad  A = - \mathcal{C}\mathfrak{C},\quad
\quad W  = \frac{1}{2} - \frac{1}{4} \left(\frac{s_0}{\ell}\right)^2 Q_3.
\end{equation}

We look for plane wave solutions of \eqref{eq:Schroedinger} of the form
\begin{equation} \label{eq:spin-wave-h}
\psi (\chi ,\tilde{t}) = \mathrm{u} e^{i\Phi } + \mathrm{v} e^{-i\Phi },\qquad \Phi  = q\chi -\Omega \tilde{t} +\eta,
\end{equation}
where $q=k s_0$  is a dimensionless wave number, $\Omega =\omega /\Omega _0$ is a dimensionless frequency, $\eta $ is an arbitrary phase, and $ \mathrm{u}, \mathrm{v}\in\mathbb{R}$ are constant amplitudes. By substituting \eqref{eq:spin-wave-h} into the generalized Schr{\"o}dinger equation \eqref{eq:Schroedinger}, we obtain 
\begin{equation} \label{eq:disp}
\fl
\Omega (q) = -2\mathcal{C}\mathfrak{C} q + \sqrt{\left[q^2 + 1 + \frac{Q_1-Q_3}{2} \left(\frac{s_0}{\ell}\right)^2 \right] \left[q^2 + \frac{Q_1}{2} \left(\frac{s_0}{\ell} \right)^2\right]},
\end{equation}
see Fig.~\ref{fig:helicoid}~(b), in which the parameters have the following values: $s_0/\ell=5$, $Q_1=0.2$, $Q_3 = -0.6 $, and $\mathcal{C} = \mathfrak{C} = 1$. The dispersion relation \eqref{eq:disp} for the helicoid ribbon is similar to that of a helical wire \cite{Sheka15c}, but different from that of a straight wire, in that it is not reflection-symmetric in $q$. The sign of the asymmetry is determined by the product of the  helicoid chirality $ \mathcal{C}$, which depends on the topology of the ribbon,  and the magnetochirality  $\mathfrak{C}$, which depends on the topology of the magnetic structure. This asymmetry stems from the curvature-induced effective Dzyaloshinskii--Moriya interaction, which is the source of the vector potential $\vec{A} = A \vec{e}_{\scriptsize \textsc{t}}$, where $A = - \mathcal{C}\mathfrak{C}$. In this context, it is instructive to mention a relation between the Dzyaloshinskii--Moriya interaction and the  Berry phase \cite{Freimuth14}.


\section{M\"obius ribbon}
\label{sec:Moebius-ribbon}

\begin{figure}
\includegraphics[width=\linewidth]{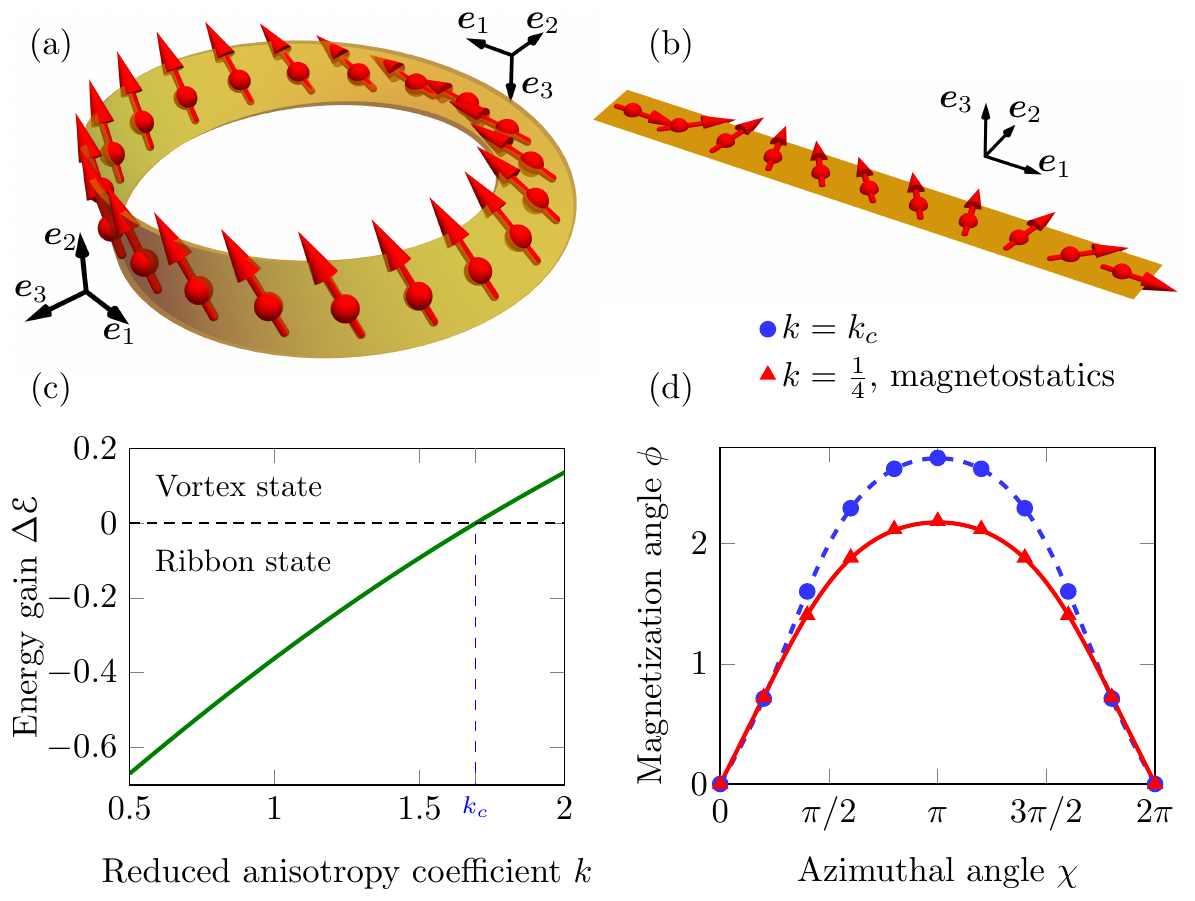}
\caption{\textbf{Magnetic M\"obius ribbon:} (a) Magnetization distribution for the ribbon state in the laboratory frame, see Eq.~\eqref{eq:rib-Four}. (b) Magnetization distribution for the ribbon state in the ribbon frame. (c)~The energy difference between the vortex and  ribbon states; when the reduced anisotropy coefficient $k$ exceeds the critical value $k_c$, see Eq.~\eqref{eq:kc}, the vortex state is favourable, while for $k<k_c$, the inhomogeneous ribbon state is realized. (d) In-surface magnetization angle $\phi$ in the ribbon state. Lines correspond to Eq.~\eqref{eq:rib-Four} and markers correspond to \textsf{SLaSi} simulations, see \ref{app:sim} for details. Red triangles represent the simulations with dipolar interaction without magnetocrystalline anisotropy ($k=0$); it corresponds very well to our theoretical result (solid red curve) for effective anisotropy $k = 1/4$ induced by magnetostatics, see Eq.~\eqref{eq:Q_1}.}
\label{fig:Moebius}
\end{figure}

In this section we consider a narrow M{\"o}bius ribbon. The  M{\"o}bius ring was studied previously in Ref.~\cite{Pylypovskyi15b}.  The ground state is determined by the relationship between geometrical and magnetic parameters. The vortex configuration is favorable in the small anisotropy case, while a topologically protected domain wall is the ground state for  large easy-normal anisotropy. Although the problem was studied for a wide range of parameters, the limit of a narrow ribbon was not considered previously. Below we show that the narrow M\"obius ribbon exhibits \emph{a new inhomogeneous ground state}, see Fig.~\ref{fig:Moebius}~(a),~(b).

The M\"obius ribbon  has a circle as its central curve and turns at a constant rate, making a half-twist once around the circle; it can be formed by joining the ends of a helicoid ribbon.
Letting $R$ denote the radius, we use the angle $\chi = s/R$ instead of arc length $s$ as parameter, and set
\begin{equation} \label{eq:circle}
\vec{\gamma}(\chi) = R\cos\chi  \hat{\vec{x}} + R\sin\chi \hat{\vec{y}}, \qquad \alpha(\chi) = \pi - \mathcal{C} \chi/2.
\end{equation}
The chirality $\mathcal{C} = \pm 1$ determines whether the M\"obius ribbon is right- or left-handed.
From \eqref{eq:surface}, the parametrized surface is given by
\begin{equation} \label{eq:Moebius}
\fl
\vec\varsigma(\chi,v)=\left(R+v\cos\frac{\chi}{2}\right) \cos\chi\, \hat{\vec x} + \left(R+v\cos\frac{\chi}{2}\right) \sin\chi \, \hat{\vec y}+\mathcal{C}v\sin\frac{\chi}{2} \, \hat{\vec z}.
\end{equation}
Here $\chi \in [0,2\pi)$ is the azimuthal angle and $v\in \left[ -w/2,w/2 \right]$ is the position along the
ring width. From \eqref{eq:E-ribbon} the energy of the narrow M\"obius ribbon reads $E = 4\pi M_s^2 hwR\int\limits_0^{2\pi} \mathscr{E} \mathrm{d}\chi$, where the energy density is given by
\begin{equation*} 
\fl
\eqalign{
\mathscr{E}&= \left(\frac{\ell}{R}\right)^2 \left[\mathcal{C} \partial_\chi \theta +\frac{1}{2}\sin\phi +  \cos\phi \sin\frac{\chi}{2}\right]^2 + \left(\frac{\ell}{R}\right)^2 
\left[\sin\theta \left(\partial_\chi \phi - \cos\frac{\chi}{2}\right)\right.\\
&\left. + \mathcal{C}\cos\theta \left(\frac12\cos\phi - \sin\phi \sin\frac{\chi}{2}\right)\right]^2 -\frac{ Q^\mathrm{eff}_1}{2}\sin^2\theta\cos^2\phi - \frac{ Q^\mathrm{eff}_3}{2}\cos^2\theta.
}
\end{equation*}
with effective anisotropies (cf \eqref{eq:Q1-Q3ribbon-2})
\begin{equation*} 
Q^{\mathrm{eff}}_1 = Q_1 - \frac{1}{2} \left(\frac{\ell}{R}\right)^2, \qquad Q^{\mathrm{eff}}_3 = Q_3 - \frac{1}{2} \left(\frac{\ell}{R}\right)^2.
\end{equation*}
The effective curvature and torsion are given by  (cf  \eqref{eq:kappa-tau})
\begin{equation*} 
\kappa^{\mathrm{eff}} = -\frac{2\cos\frac{\chi}{2}}{R}\frac{1 + 2\sin^2\frac{\chi}{2}}{1 + 4 \sin^2\frac{\chi}{2}}, \qquad \tau^{\mathrm{eff}} = -\frac{\mathcal{C}}{2R}\sqrt{1+4\sin^2\frac{\chi}{2}}.\\
\end{equation*}

Let us consider the case of uniaxial magnetic materials, for which $K_3 = 0$.  Under the reasonable assumption  $\ell \ll R$,  we have that $Q_3^{\mathrm{eff}}\approx -1$, so that the easy-surface anisotropy dominates the energy density and acts as an in-surface constraint (as for the helicoid ribbon).  Taking $\theta = \pi/2$, we obtain the simplified energy density
\begin{equation*} 
\fl
\eqalign{
\mathscr{E} &= \left(\frac{\ell}{R}\right)^2 \left(\partial_\chi \phi - \cos\frac{\chi}{2}\right)^2 + \left(\frac{\ell}{R}\right)^2 \left(\frac{1}{2}\sin\phi +  \cos\phi \sin\frac{\chi}{2}\right)^2  -\frac{Q_1^{\mathrm{eff}}}{2}\cos^2\phi,
}
\end{equation*}
which depends only on the in-surface  orientation $\phi$. The equilibrium magnetization distribution is described by the following Euler--Lagrange equation  
\begin{equation} \label{eq:Moebius-EL-phi}
\eqalign{
&\partial_{\chi\chi} \phi + \sin\frac{\chi}{2}\,\sin^2\!\phi+  \left(\sin^2\frac{\chi}{2} - k\right) \sin \phi \cos\phi = 0,\\
&\phi(0)=-\phi(2\pi)\; \mathop{\rm mod} 2\pi, \qquad \partial_\chi \phi(0)=-\partial_\chi\phi(2\pi), }
\end{equation}
where the antiperiodic boundary conditions compensate for the half-twist in the M\"obius ribbon and ensure that the magnetisation $\vec{m}$ is smooth at $\chi = 0$. The reduced anisotropy coefficient $k$ in \eqref{eq:Moebius-EL-phi} reads
\begin{equation} \label{eq:k-reduces}
k =  \frac{Q^{\mathrm{eff}}_1}{2}\left(\frac{R}{\ell} \right)^2 = \frac{K_1}{4\pi M_s^2}\frac{R^2}{\ell^2} + \frac{h R^2}{2 \pi w \ell^2} \ln\frac{w}{h} - \frac{1}{4}.
\end{equation}
It is easily seen that if $\phi(\chi)$ is a solution of \eqref{eq:Moebius-EL-phi}, then $\phi(\chi) + n\pi$ is also a solution with the same energy.  Since solutions differing by multiples of $2\pi$ describe the same magnetisation, only $\phi(\chi) +\pi$ corresponds to a configuration distinct from $\phi$.  We also note that if $\phi(\chi)$ is a solution of \eqref{eq:Moebius-EL-phi}, then $-\phi(-\chi)$ is also a solution  with the same energy.

By inspection, $\phi_+^{\rm vor} \equiv 0$ and $\phi_-^{\rm vor} \equiv \pi$,  are solutions of \eqref{eq:Moebius-EL-phi}; the ground states are
\begin{equation*} 
\theta^{\rm vor} = \frac{\pi}{2}, \qquad \cos \phi^{\rm vor} = \mathfrak{C},
\end{equation*}
where the magnetochirality  $\mathfrak{C}$ determines whether the magnetisation $\vec{m}$ is parallel or antiparallel to the circular axis.
We refer to these as  {\it vortex states}.  Unlike the case of the helicoid ribbon, $\phi \equiv \pm \pi/2$ is not a solution of \eqref{eq:Moebius-EL-phi}.  Numerically, we find two further  solutions  of the Euler-Lagrange equation, denoted $\phi_+^{\rm rib}(\chi)$ and $\phi_-^{\rm rib}(\chi) \equiv \phi_+^{\rm rib}(\chi) + \pi$, which we call \emph{ribbon states}. While we have not obtained analytical expressions for $\phi^{\rm rib}(\chi)$,  good approximations can be found by assuming $\phi_+^{\rm rib}$ to be antiperiodic and odd, so that it has a Fourier-sine expansion of the form
\begin{equation} \label{eq:rib-Four}
\phi_+^{\rm rib}(\chi)  = \sum_{n=1}^\infty c_n \sin((2n-1)\chi/2).
\end{equation}
The series is rapidly converging, with the  first four coefficients $c_1 = 2.245$, $c_2 = 0.0520$, $c_3 = -0.0360$, and $c_4 =-0.0142$ for $k = 0.25$, providing an approximation accurate to within 0.03\% (specifically, the $L^2$-norm difference between the numerically determined $\phi$, as described in \ref{app:sim}, and this expansion is 0.003).

Numerical calculations indicate that the ground state of the M\"obius ribbon, like the helicoid ribbon,  undergoes a bifurcation as the tangential-axis anisotropy decreases.  Unlike the helicoid ribbon, the bifurcation occurs for positive anisotropy $k_c$ given by
\begin{equation} \label{eq:kc}
k_c\approx 1.6934 .\end{equation}
For $k>k_c$, the vortex state has the lowest energy,  whereas for  $k<k_c$,  the ribbon state has the lowest energy.  
The energy difference between the vortex and ribbon states,
\begin{equation*} 
\Delta \mathcal{E} = \frac{E^{\mathrm{rib}} - E^{\mathrm{vor}}}{2 M_s^2 h w R} = \frac{1}{2\pi} \int\limits_0^{2\pi} \mathcal{E} \mathrm{d} \chi - \frac{5}{4},
\end{equation*}
is plotted in Fig.~\ref{fig:Moebius}(c). In some respects the ribbon state resembles an onion state in magnetic rings  \cite{Klaui03a, Kravchuk07, Guimaraes09, Sheka15}; in the laboratory reference frame the magnetization distribution is close to a spatially homogeneous state, see Fig.~\ref{fig:Moebius}(a).

The  in-surface magnetization angle  $\phi(\chi) $ for the ribbon state is plotted in  Fig.~\ref{fig:Moebius}(d). The plot shows good agreement between the  analytic expression \eqref{eq:rib-Four} and  spin--lattice \textsf{SLaSi} simulations (see \ref{app:sim} for details). The blue dashed line with solid circles represents the case $k=k_c$ (the critical anisotropy value). The red solid line corresponds to the solution of Eq.~\eqref{eq:Moebius-EL-phi} for $k = 1/4$, an effective anisotropy induced by magnetostatics. It is in a good agreement with simulations shown by red triangles where the dipole--dipole interaction is taken into account instead of easy-tangential anisotropy. The magnetization distribution for the ribbon state is shown in Fig.~\ref{fig:Moebius}(a) (3-dimensional view) and Fig.~\ref{fig:Moebius}(b) (an untwisted schematic of the M\"{o}bius ribbon).

Let us estimate the values of the parameters for which the ribbon state is energetically preferable. Taking into account \eqref{eq:k-reduces}, we find that the ribbon state is energetically preferable provided
\begin{equation*} 
\frac{K_1}{4\pi M_S^2} < \left(k_c + \frac{1}{4}\right) \frac{\ell^2}{R^2} - \frac{h}{2 \pi w} \ln\frac{w}{h}.
\end{equation*}
This condition is \emph{a fortiori} satisfied for the hard axial case, ie when $K_1<0$. For soft magnetic materials $(K_1=0)$ the only source of anisotropy is the shape anisotropy. The ribbon state is the ground state when
\begin{equation*} 
\frac{h R^2}{2 \pi w \ell^2} \ln\frac{w}{h} <k_c + \frac{1}{4},
\end{equation*}
which imposes constraints on the geometry and material parameters. 


\section{Conclusion}
\label{sec:conclusion}

We have studied ferromagnetic ribbons, that is magnetic materials in the shape of thin shells whose median surface is swept out by a line segment turning as it moves along a central curve. Ferromagnetic ribbons combine properties of both 1D systems, ie nanowires, and 2D systems, ie curved films and nanoshells. While the geometrical properties of a narrow ribbon are described by its central curve and the rate of turning of its transverse line segment, its magnetic properties are determined by the geometrical and magnetic properties of the ribbon surface. The micromagnetic energy of the ribbon can be reduced to the energy of a 1D system (magnetic nanowire) with effective curvature, torsion and biaxial anisotropy. While the source of effective curvature and torsion is the exchange interaction only, the biaxiality results from both exchange and magnetostatics.

We have studied two examples: (i)  a narrow  helicoid ribbon and ii) a narrow M\"obius ribbon. The helicoid ribbon has zero effective curvature but finite torsion, which provides a paradigmatic model for studying \emph{purely torsion-induced effects}. Similar to a microhelix structure \cite{Sheka15c}, a geometry-induced effective Dzyaloshinskii--Moriya interaction is a source of coupling between the helicoid chirality and the magnetochirality, which essentially influences both magnetization statics and dynamics.  The emergent magnetic field  generated by the torsion breaks  mirror symmetry, so that the properties of magnetic excitations in different spatial directions is not identical.
 The narrow M\"obius ribbon is characterized by  spatially varying effective curvature and torsion. We have predicted a new inhomogeneous \emph{ribbon state} for the M\"obius ribbon, which is characterized by an inhomogeneous in-surface magnetization distribution. The existence of this state has been confirmed by  spin--lattice simulations.

\ack

A.~G. acknowledges the support of EPSRC Grant No. EP/K024116/1.V.~P.~K. acknowledges the Alexander von Humboldt Foundation for the support and IFW Dresden for kind hospitality. D.~D.~Sh. thanks the University of Bristol, where part of this work was performed, for kind hospitality. J.~M.~R. and V.~S. acknowledge the support of EPSRC Grant No. EP/K02390X/1. 

\appendix


\section{Magnetostatic energy of ribbons and strips}
\label{app:rigorous}

Here we justify formulae \eqref{eq:Q1-Q3ribbon-1}, \eqref{eq:Q1-Q3ribbon-2} for the magnetostatic energy of a narrow ribbon. To this end, we calculate the magnetostatic energy of the shell of reduced width $\widetilde{w} = w/\ell$ and reduced thickness $\tilde h=h/\ell$ in the regime 
\begin{equation*} 
{\widetilde{w}}^2\leq \tilde h \leq \widetilde{w} \ll 1
\end{equation*}
and then identify the leading contributions to the energy of a ribbon in the  limit of small aspect ratio
\begin{equation*} 
\delta \equiv \frac{\tilde h}{\widetilde{w}} = \frac{h}{w}\ll 1 \,.
\end{equation*} 
For the sake of clarity, before turning our attention to the general case, we first consider a flat strip $V_s=[0,L] \times [-\frac{w}{2},\frac{w}{2}] \times [-\frac{h}{2},\frac{h}{2}]$. The magnetostatic energy may be written in the form 
\begin{equation*} 
\fl
E_{\mathrm{ms}}^{\mathrm{strip}} = - \frac{M_S^2}{2} \int\limits_V \mathrm{d} \vec{r} \int\limits_V \mathrm{d} \vec{r'} \left(\vec{m}(\vec{r}) \cdot \vec{\nabla}\right) \left(\vec{m}(\vec{r'}) \cdot \vec{\nabla'}\right)\frac{1}{\left|\vec{r} - \vec{r'}\right|}.
\end{equation*}

It is well known that the leading order contribution to the magnetostatic energy is coming from the interaction between the surface charges of the largest surfaces.  We denote by $T$ and $B$ the pair of top and bottom surfaces of the strip (of surface area $Lw$) and by $F$, $R$ the front and rear surfaces of the strip (of surface area $Lh$), respectively. It is straightforward to show (see e.g. \cite{Kohn05a, Slastikov12}) that
\begin{equation*} 
\fl
\eqalign{
\frac{2}{M_S^2} E_{\mathrm{ms}}^{\mathrm{strip}} &= \int\limits_{T \cup B} \!\! \mathrm{d}S \!\! \int\limits_{T \cup B} \!\! \mathrm{d}S' \frac{\left(  \vec{\bar m}(\vec{r}) \cdot \vec{n} \right) \left(  \vec{\bar m}(\vec{r'}) \cdot \vec{n}' \right)}{|\vec{r}-\vec{r'}|}\\
& + \int\limits_{F \cup R} \!\! \mathrm{d}S \!\! \int\limits_{F \cup R} \!\! \mathrm{d}S' \frac{\left(  \vec{\bar m}(\vec{r}) \cdot \vec{n} \right) \left(  \vec{\bar m}(\vec{r'}) \cdot \vec{n}' \right)}{|\vec{r}-\vec{r'}|} + \mathcal{O} \left(\widetilde{w}^2 \tilde{h}\right)\\
&= 2 \int\limits_0^L \!\! \mathrm{d}s  \! \int\limits_0^L \!\! \mathrm{d}s' \!\!\!\! \int\limits_{-w/2}^{w/2} \!\!\! \mathrm{d}u \!\!\!\! \int\limits_{-w/2}^{w/2} \!\!\! \mathrm{d}v \left[ \frac{ \bar m_3(s) \bar m_3(s')}{\rho} -\frac{ \bar m_3(s) \bar m_3(s')}{\sqrt{\rho^2+h^2}} \right] \\
&+ 2 \int\limits_0^L \!\! \mathrm{d}s  \! \int\limits_0^L \!\! \mathrm{d}s' \!\!\!\! \int\limits_{-h/2}^{h/2} \!\!\! \mathrm{d}u \!\!\!\! \int\limits_{-h/2}^{h/2} \!\!\! \mathrm{d}v \left[ \frac{ \bar m_2(s) \bar m_2(s')}{\rho} -\frac{ \bar m_2(s) \bar m_2(s')}{\sqrt{\rho^2+h^2}} \right]  + \mathcal{O} \left(\widetilde{w}^2 \tilde{h}\right),
}
\end{equation*}
where $\vec{\bar m}(s) = \frac{1}{wh} \int \vec{m}(s,u,v)\,\mathrm{d}u\, \mathrm{d}v$ is the average of magnetization $\vec{m}$ over the cross-section of area $wh$, $\vec{n}$ is the surface normal, and $\rho=\sqrt{\left(s-s'\right)^2+(u-v)^2}$.

We note that for an arbitrary smooth function $f$ and a constant $a$ 
\begin{equation*} 
\fl
\eqalign{
\int\limits_0^L \frac{f(s')\mathrm{d}s'}{\sqrt{a^2+\left(s-s'\right)^2}}  &= f(L) \ln \left(L-s+\sqrt{(L-s)^2 +a^2} \right)+f(0) \ln \left(s+\sqrt{s^2 +a^2} \right)\\
&-2f(s)\ln |a| - \int\limits_s^L f^\prime(s')\ln\left( \left|s-s'\right| + \sqrt{\left(s-s'\right)^2+a^2} \right) \,\mathrm{d}s' \\
& +\int\limits_0^s f^\prime(s') \ln\left(\left|s-s'\right|+\sqrt{\left(s-s'\right)^2+a^2}\right) \, \mathrm{d}s'.
}
\end{equation*}
Applying this formula and following the approach developed in \cite{Slastikov12}, we can show that the main contribution to the magnetostatic energy will be coming from the term $-2f(s)\ln |a|$ in the last integral. Therefore, we obtain
\begin{equation} \label{eq:en-flat}
\fl
\eqalign{
\frac{ E_{\mathrm{ms}}^{\mathrm{strip}} }{2 M_S^2}  &=  w^2 
\int\limits_0^L \!\! \mathrm{d}s \!\!\!\! \int\limits_{-1/2}^{1/2} \!\!\!\! \mathrm{d}u \!\!\!\!\int\limits_{-1/2}^{1/2} \!\!\!\! \mathrm{d}v\, \bar m_3^2(s) \left[ \ln \sqrt{(u-v)^2 + \delta^2} - \ln |u-v|  \right]\\
&+ h^2 \int\limits_0^L \!\! \mathrm{d}s \!\!\!\! \int\limits_{-1/2}^{1/2} \!\!\!\! \mathrm{d}u \!\!\!\!\int\limits_{-1/2}^{1/2} \!\!\!\! \mathrm{d}v\, \bar m_2^2(s) \left[ \ln \sqrt{(u-v)^2 + 1/\delta^2} - \ln |u-v|  \right] + \mathcal{O}\left( \widetilde{w} \bar{h}^2 \left|\ln \bar{h}\right|\right).  
} 
\end{equation}
By integrating over the cross-section variables, the expression \eqref{eq:en-flat} further simplifies to
\begin{equation*} 
\fl
\eqalign{
\frac{ E_{\mathrm{ms}}^{\mathrm{strip}} }{2 M_S^2}  &= wh \left( 2 \arctan \frac1
{\delta} + \delta \ln \delta+ \left(\frac{1}{2 \delta} - \frac{\delta}{2 } \right) \ln (1+\delta^2) \right) \int\limits_0^L \bar m_3^2(s)  \, \mathrm{d}s  \\
&+ wh \left( - \delta \ln \delta + \frac{2}{\delta} \arctan \delta + \left(\frac{\delta}{2}- \frac{1}{\delta} \right) \ln (1+\delta^2)  \right) \int\limits_0^L \bar m_2^2(s)  \, \mathrm{d}s.
} 
\end{equation*}
Hence, the magnetostatic energy of the flat strip is 
\begin{equation} \label{eq:en-flat-2}
\fl
\eqalign{
E_{\mathrm{ms}}^{\mathrm{strip}} = {2\pi M_S^2 hw} \left\{ \int\limits_0^L \left[ \left(1+\frac{\delta}{\pi} \ln \delta\right) \bar m_3^2(s)- \frac{\delta}{\pi}   \ln \delta \, \bar m_2^2(s) \right] \mathrm{d}s + \mathcal{O}(\delta) \right\} \,.
} 
\end{equation}

Returning to the general case, we recall from \eqref{eq:surface} that a ribbon may be parametrized as 
\begin{equation*}
\vec{\varsigma}(s,v)=\vec{\gamma}(s)+v \vec{e_2}(s) \,, \qquad v \in \left[-\frac{w}{2},\frac{w}{2} \right] , \; s \in \left[0,L\right]
\end{equation*}
and consider a shell of thickness $h$ around $\vec{\varsigma}$ parametrized as 
\begin{equation*}
\vec{\varrho}(s,v,u)=\vec{\gamma}(s)+v\vec{e_2}(s)+u \vec{e_3}(s,v) \,,
\end{equation*}
where $\vec{e_2}, \vec{e_3}$ are defined in \eqref{eq:2D-basis} and $h$ is small enough so that $\vec{\varrho}$ does not intersect itself. Then, introducing $\bar{m}_2 = \bar{\vec{m}} \cdot \vec{e}_2$ and $\bar{m}_3 = \bar{\vec{m}} \cdot \vec{e}_3$, the energy of the shell up to terms of order $\mathcal{O}\left( \widetilde{w} \bar{h}^2 \left|\ln \bar{h}\right|\right)$ is given by 
\begin{equation} \label{eq:en-ribshell}
\fl
\eqalign{
\frac{ E_{\mathrm{ms}}^{\mathrm{ribbon}} }{2 M_S^2}  &=  w^2 
\int\limits_0^L \!\! \mathrm{d}s \!\!\!\! \int\limits_{-1/2}^{1/2} \!\!\!\! \mathrm{d}u \!\!\!\!\int\limits_{-1/2}^{1/2} \!\!\!\! \mathrm{d}v\, \sqrt{g} \bar{m}_3(s,wu) \bar{m}_3(s,wv) \ln \frac{\sqrt{(u-v)^2 + \delta^2}}{|u-v|} \\
& +h^2 \int\limits_0^L \!\! \mathrm{d}s \!\!\!\! \int\limits_{-1/2}^{1/2} \!\!\!\! \mathrm{d}u \!\!\!\!\int\limits_{-1/2}^{1/2} \!\!\!\! \mathrm{d}v\, \bar m_2^2(s) \left[ \ln \sqrt{(u-v)^2 + 1/\delta^2} - \ln |u-v|  \right].
} 
\end{equation}

We remark that the formula \eqref{eq:en-ribshell} yields the correct result both for a wire with a rectangular cross-section ($h/w=\mathrm{const}$) and in the thin film limit ($h/w \rightarrow 0$), cf. \cite{Slastikov12} and \cite{Slastikov05}, respectively, however in the latter case it resolves  terms beyond the leading order. 

Expanding the first integral in \eqref{eq:en-ribshell} in $w$ and integrating over cross-section variables, we obtain that the magnetostatic energy of the ribbon 
\begin{equation} \label{eq:en-rib}
\fl
\eqalign{
E_{\mathrm{ms}}^{\mathrm{ribbon}} = {2\pi M_S^2 hw} \left\{ \int\limits_0^L \left[ \left(1+\frac{\delta}{\pi} \ln \delta\right) \bar{m}_3^2(s) - \frac{\delta}{\pi}  \ln \delta \, \bar{m}_2^2(s) \right] \mathrm{d}s + \mathcal{O}(\delta) \right\}
} 
\end{equation}
is insensitive to curvature effects, cf \eqref{eq:en-flat-2}. Finally, using the constrain $\vec{\bar{m}}^2=1$ we get the magnetostatic energy in the form \eqref{eq:Q1-Q3ribbon-1}, \eqref{eq:Q1-Q3ribbon-2}.


\section{Simulations}
\label{app:sim}

We use the in-house developed spin-lattice simulator \textsf{SLaSi} \cite{SLaSi}. A chain of classical magnetic moments $\vec{m}_i$, $|\vec{m}_i| = 1$, $i =\overline{1,N}$ is considered. They are situated on a circle \eqref{eq:circle}, which defines a central axis of the narrow M\"{o}bius ribbon \eqref{eq:Moebius}, hence the periodic condition $\vec{m}_{N+1} = \vec{m}_1$ is used. The following classical Hamiltonian is used:
\begin{equation} \label{eq:hamiltonian}
\fl
\eqalign{
\mathscr{H} & = - a\ell^2 \sum_{i = 1}^{N} (\vec{m}_{i}\cdot \vec{m}_{i+1}) - \frac{a^3}{2}\sum_{i = 1}^{N} \left[ Q_1 (\vec{m}_i\cdot \vec{e}_{1i})^2 +  Q_3 (\vec{m}_i\cdot \vec{e}_{3i})^2 \right] \\
 & + d \frac{a^3}{8\pi} \sum_{i\neq j} \left[ \frac{(\vec{m}_i\cdot \vec{m}_j)}{r_{ij}^3} - 3 \frac{(\vec{m}_i\cdot\vec{r}_{ij})( \vec{m}_j\cdot\vec{r}_{ij})}{r_{ij}^5} \right],
}
\end{equation}
where $a$ is the lattice constant, $\vec{e}_{1i}$ and $\vec{e}_{3i}$ are unit basis vectors \eqref{eq:2D-basis} in $i$-th site and the coefficient $d = 0, 1$ is used as a switch for dipolar interactions. 

To study the static magnetization distribution, we minimize the energy by solving a set of $N$ vector  Landau--Lifshitz--Gilbert ordinary differential equations for $N = 100 $ sites situated on a ring of radius $R = aN/(2\pi)$ and $\ell = R$ using the Runge--Kutta--Fehlberg scheme (RKF45), see~\cite{Pylypovskyi14} for general description of the simulator. The equilibrium magnetization state is found starting the simulations from different initial distributions (four different random ones, uniformly magnetized states along $\pm \hat{\vec{x}}$, $\pm \hat{\vec{y}}$, $\pm \hat{\vec{z}}$ and along unit vectors $\vec{e}_i$.

The simulations are performed using the high-performance computer clusters of the Taras Shevchenko National University of Kyiv~\cite{unicc} and the Bayreuth University~\cite{btrzx}.

%
%
%
%
\providecommand{\newblock}{}

%
\end{document}